\documentclass[a4paper]{panl}
\usepackage{cite}
\usepackage{wrapfig}
\usepackage{graphicx}
\usepackage{amssymb}
\usepackage{amsfonts}
\usepackage{amsmath}
\usepackage{longtable}
\usepackage{rotating}
\usepackage{lscape}
\usepackage{epsfig}
\usepackage{multirow}
\originalTeX
\def\bfg #1{{\mbox{\boldmath $#1$}}}
\begin{document}
\issuearea{Physics of Elementary Particles and Atomic Nuclei. Theory}

\title{Spin observables of proton-deuteron elastic scattering at SPD NICA energies within the Glauber model and pN amplitudes}

\maketitle
\authors{
Yu.\,Uzikov$^{\,a,b,c}$\footnote{E-mail: uzikov@jinr.ru} , 
J.\,Haidenbauer$^{\,d,}$\footnote{E-mail: j.haidenbauer@fz-juelich.de},
A. Bazarova$^{\,e,}$\footnote{E-mail: alba.9@mail.ru}
A.\,Temerbayev$^{\,e,f}$\footnote{E-mail:adastra77@mail.ru}}
\from{$^{a}$\,Joint Institute for Nuclear Researches, Dubna, Moscow reg. 141980 Russia}
\vspace{-3mm}
\from{$^{b}$\,Dubna State University, Dubna, Moscow reg. 141980 Russia }
\vspace{-3mm}
\from{$^{c}$\, Department of Physics, M.V. Lomonosov State University, Moscow, 119991 Russia }
\vspace{-3mm}
\from{$^{d}$\,Institute for Advanced Simulation and Institut f\"ur Kernphysik, Forschungszentrum J{\"u}lich GmbH, D-52425 J\"ulich, Germany}
\vspace{-3mm}
\from{$^{e}$\,L.N. Gumilyov Eurasian National University, Nur-Sultan, 010000, Kazakhstan}
\from{$^{f}$\,Institute of Nuclear Physics, Astana branch, Nur-Sultan, 010000, Kazakhstan}
\begin{abstract}
\vspace{0.2cm}
A systematic analysis of nucleon-nucleon scattering amplitudes is available up to a laboratory energy of 
$3$~GeV in case of the $pp$ system and up to $1.2$ GeV for $pn$. 
At higher energies there is only incomplete experimental information
on $pp$ elastic scattering, whereas data for the $pn$ system are very scarce.
We apply the spin-dependent Glauber theory to calculate spin observables of $pd$ elastic scattering at $3$-$50$ GeV/c using $pp$ amplitudes available in the literature
and parametrized
within the Regge formalism. The calculated vector $A_y^p$, $A_y^d$ 
and tensor $A_{xx}$, $A_{yy}$ analyzing powers  
 and the spin-correlation coefficients $C_{y,y}$, $C_{x,x}$,
$C_{yy,y}$, $C_{xx,y}$ can be measured at 
SPD NICA and, thus, will provide a test of the used $pN$ amplitudes. 
\end{abstract}
\vspace*{6pt}


\section*{Introduction}
\label{sec:intro}

The spin amplitudes of $pp$ and $pn$ elastic scattering contain important information on the dynamics of the $NN$ interaction. 
A systematic reconstruction of these amplitudes from scattering data is provided by the SAID partial-wave analysis \cite{Arndt:2007qn} 
and covers laboratory energies up to $3$ GeV ($p_{lab}\approx 3.8$~GeV/c) for $pp$ and $1.2$ GeV ($p_{lab}\approx 1.9$~GeV/c)
for $pn$ scattering. At higher energies there is only incomplete experimental information on $pp$ scattering, whereas data
for the $pn$ system are very scarse. In the literature there are some parametrizations for $pN$ amplitudes, 
obtained in the eikonal model \cite{Sawamoto:1979cb} for the laboratory momentum $6$ GeV/c and within the Regge 
phenomenology \cite{Sibirtsev:2009cz} for $3$-$50$ GeV/c (corresponding to $2.77 < \sqrt{s} < 10$ GeV).
Another Regge-type parametrization for values of $s$ above $6$ GeV$^2$ ($p_{lab} \ge 2.2$~GeV/c)
was presented in Ref.~\cite{Ford:2012dg}. 
A possible way to check existing parametrizations is to study spin effects in proton-deuteron ($pd$) and neutron-deuteron ($nd$) 
elastic and quasi-elastic scattering. At high energies and small four-momentum transfer $t$, $pd$ scattering can be described 
by the Glauber diffraction theory of multistep scattering, which involves as input on-shell $pN$ elastic scattering amplitudes. 
Applications of this theory with spin-dependent effects 
included \cite{Platonova:2016xjq} indicate a good agreement with 
the $pd$ scattering data at energies about $1$~GeV if the SAID values for the $pN$ scattering amplitudes are used 
as starting point of the calculations 
\cite{Temerbayev:2015foa,Mchedlishvili:2018uur,Platonova:2019yzf}.

In the present work we apply the spin-dependent Glauber theory 
\cite{Platonova:2016xjq,Temerbayev:2015foa} to calculate spin observables of $pd$ 
elastic scattering at $5$-$50$ GeV/c utilizing the $pp$ elastic scattering amplitudes 
established and parametrized in Ref.~\cite{Sibirtsev:2009cz} within  
the Regge formalism. As a first approximation, for the $pn$ amplitudes we use likewise the ones for $pp$ 
from \cite{Sibirtsev:2009cz}. We should note that, in principle, the Regge approach allows one to construct $pn$ 
(and $\bar p N$) amplitudes together with the $pp$ amplitudes. 
However, in view of the scarce experimental information on the spin-dependent $pn$ amplitudes and taking 
into account that the spin-independent parts of the $pp$ and $pn$ amplitudes at high energies 
are approximately the same, we assume here that the whole $pn$ amplitude is the same as that for $pp$.
The calculated vector $A_y^p$, $A_y^d$ and tensor analyzing powers $A_{xx}$, $A_{yy}$, 
and the spin-correlation coefficients $C_{y,y}$, $C_{x,x}$,
$C_{yy,y}$, $C_{xx,y}$ can be measured at SPD NICA \cite{Savin:2015paa}, 
which will provide a serious test of the used $pN$ amplitudes. 
A knowledge of  $pN$ helicity amplitudes is required in preparation and subsequent 
analysis of experiments for search of time-reversal 
invariance violation effects in double-polarized $pd$ scattering \cite{Uzikov:2015aua,Uzikov:2016lsc}.


\section*{Elements of formalism}
\label{sec:preparation}

The reaction amplitude $pd\to pd$ can be written as \cite{Temerbayev:2015foa}
\begin{eqnarray}
  <p'\mu',d'\lambda'|T|p\mu,d\lambda>= 
\varphi^+_{\mu'}{e_\beta^{(\lambda')}}^*T_{\beta \alpha} ({\bf p}, {\bf p}',\bfg\sigma)
e_\alpha^{(\lambda)}\varphi_{\mu},
\label{tfi}
\end{eqnarray}
where
$\varphi_{\mu}$ ($\varphi_{\mu'}$) is the Pauli spinor of the initial (final) proton
in the state with spin projecion $\mu$ ($\mu'$) onto the quantization axis, 
$e_\alpha^{(\lambda)}$ ($e_\beta^{(\lambda')}$) is the polarization vector
of the initial $d$ (final $d'$) deuteron in the state with the spin projection
 $\lambda$ ($\lambda'$), and
 $T_{\beta \alpha}$ is the dynamical tensor ($\beta, \alpha$=$x,y,z$) acting 
in the spin space of the proton.
The latter depends on the momentum  of the initial ($\bf p$) and final (${\bf p}'$)
proton and contains the Pauli spin matrices
 ${\bfg \sigma}=(\sigma_x, \ \sigma_y, \ \sigma_z)$.
  We use the Madison reference frame with the axis OZ$||$${\bf p}$, OY$||$$ [{\bf p}\times {\bf p}']$
 and  OX choosen in such a way to provide a right-handed coordinate system.
Using the method of Ref.~\cite{Uzikov:1998qk} for the transition tensor
$T_{\beta \alpha}$ and assuming parity (P) conservation, one has, in the general case, 
independently of the dynamics of the considered process, 
\begin{eqnarray} 
\begin{array}{ccc}
T_{xx}=M_1+M_2\sigma_y & T_{xy}=M_7\sigma_z+M_8\sigma_x  &T_{xz}=M_9+M_{10}\sigma_y  \\
T_{yx}=M_{13}\sigma_z+M_{14}\sigma_x& T_{yy}=M_3+M_4\sigma_y  & T_{yz}=M_{11}\sigma_x+M_{12}\sigma_z   \\
T_{zx}=M_{15}+M_{16}\sigma_y & T_{zy}=M_{17}\sigma_x+M_{18}\sigma_z  & T_{zz}=M_5+M_6\sigma_y,
\end{array}
\label{Txx-Tzz}
\end{eqnarray}
 where
$M_i$ ($i=1,\dots, 18$) are complex amplitudes determined by the dynamics of the reaction. 
As was mentioned in Ref. \cite{Temerbayev:2015foa}, under the operation of time reversal this reference
 frame is rotated around the OY axis on the scattering angle $\theta$ that is the angle between the vectors 
$\bf p$  and $\bf p^\prime$. This complicates the formulation of the time-reversal (T) invariance conditions 
as compared to other (non-Madison) reference frames, like in Ref.~\cite{Platonova:2016xjq} and, 
therefore, we do not write them explicitly. However, the amplitudes $M_i$ ($i=1,\dots,18$) satisfy T-reversal 
invariance since they are explicitly expressed as linear combinations of the T-reversal invariant amplitudes 
$A_i$ ($i=1,\dots,12$) of $pd$ scattering introduced in Ref.~\cite{Platonova:2016xjq}. 
 
The spin observables $A_y$, $A_{ij}$, and $C_{ij,k}$ defined in the
notation of Ref.~\cite{Ohlsen:1972zz} and considered in the present work,
have the following form in terms of the amplitudes $M_i$
 \begin{eqnarray}
 \label{observs}
\frac{d\sigma}{dt}&=&\frac{1}{6}Tr MM^+,\ \ \ \ \ \
Tr MM^+=2\sum_{i=1}^{18}|M_i|^2,\\ \nonumber
A_y^d&=&Tr M S_yM^+/TrMM^+ = \\ \nonumber 
&-&\frac{2}{\sum_{i=1}^{18}|M_i|^2} Im(M_1M_9^*+M_2M_{10}^*+M_{13}M_{12}^*+M_{14}M_{11}^*
+M_{15}M_5^*+M_{16}M_6^*),\\ \nonumber 
A_y^p&=&TrM \sigma_yM^+/TrMM^+= \\ \nonumber 
&&\frac{2}{\sum_{i=1}^{18}|M_i|^2} \Bigl [Re(M_1M_2^*+M_9M_{10}^*+M_{3}M_{4}^*+M_{15}M_{16}^*
+M_{5}M_6^*) -\\ \nonumber
&&Im(M_8M_7^*+M_{14}M_{13}^*+ M_{11}M_{12}^*+M_{17}M_{18}^*)\Bigr ], \\ \nonumber
A_{yy}&=&TrM {\cal P}_{yy} M^+/TrMM^+= \\ \nonumber
&& 1-\frac{ 3}{\sum_{i=1}^{18}}(|M_3|^2 +|M_4|^2+|M_7|^2+|M_8|^2+M_{17}|^2+|M_{18}|^2),\\ \nonumber
 A_{xx}&=&TrM {\cal P}_{yy} M^+/TrMM^+= \\ \nonumber
&& 1-\frac{ 3}{\sum_{i=1}^{18}}(|M_1|^2 +|M_2|^2+|M_{13}|^2+|M_{14}|^2+|M_{15}|^2+|M_{16}|^2),\\ \nonumber 
\end{eqnarray}
\begin{eqnarray}
\label{observs1}
\nonumber
  C_{y,y}&=&TrM{ S}_y\sigma_yM^+/TrMM^+= \\ \nonumber 
&&-\frac{2}{\sum_{i=1}^{18}|M_i|^2}\Bigl [Im(M_2M_9^*+M_1M_{10}^*
+M_{16}M_{5}^*+M_{15}M_6^*)+\\ \nonumber
&& Re(M_{14}M_{12}^* -M_{13}M_{11}^*)\Bigr ], \\ \nonumber
 C_{x,x}&=&TrM{ S}_y\sigma_yM^+/TrMM^+= \\ \nonumber 
&&-\frac{2}{\sum_{i=1}^{18}|M_i|^2}\bigl [Im(M_8M_9^*+M_3M_{11}^*+M_{17}M_{5}^*)+\\ \nonumber
&& Re(M_{7}M_{10}^* -M_{4}M_{12}^* +M_{18}M_{6}^*)\Bigr ], \\ \nonumber 
  C_{xx,y}&=&TrM{\cal }_{xx}\sigma_yM^+/TrMM^+= \\ \nonumber 
&&=A_y^p-\frac{6}{\sum_{i=1}^{18}|M_i|^2}
\frac{2}{\sum_{i=1}^{18}|M_i|^2}\bigl [Re (M_2M_1^*+M_{16}M_{15}^*)- Im(M_{14}M_{13}^*)\\ \nonumber
  C_{yy,y}&=&TrM{\cal }_{yy}\sigma_yM^+/TrMM^+= \\ 
&-&=A_y^p-\frac{6}{\sum_{i=1}^{18}|M_i|^2}
\bigl [ReM_{3}M_{4}^*-Im(M_{17}M_{18}^*+M_{8}M_{7}^*) \Bigr ], \nonumber
\end{eqnarray}
where ${\cal P}_{ij}=\frac{3}{2}(S_iS_j+S_jS_i)-2\delta_{ij}$ and $S_j$ ($j=x,y,z$) are Cartesian 
components of the spin operator for the system with $S=1$.

In Ref.~\cite{Platonova:2016xjq} (PK) the general spin structure of the 
transition operator $pd\to pd$ is written in a different representation and 
in another reference frame with axis $OX' \uparrow\uparrow {\bf q}$, 
$OZ' \uparrow\uparrow {\bf k}$, $OY' \uparrow\uparrow {\bf n}$,
where ${\hat {\bf q}}=({\bf p}-{\bf p}')/|({\bf p}-{\bf p}')|$, 
${\hat {\bf k}}= ({\bf p}+{\bf p}')/|({\bf p}+{\bf p}')|$, ${\hat {\bf n}}= [\bf{k}\times {\bf q}]$ forming
the right-handed system (${\bf p}$ and ${\bf p}'$ being the momenta of the incident and outgoing proton,
respectively).

The amplitudes of $pN$ elastic scattering are written as \cite{Platonova:2016xjq}
\begin{eqnarray}
\label{pnamp}
M_N({\bf p}, {\bf q};\bfg \sigma, {\bfg \sigma}_N)=
 A_N+C_N\bfg \sigma \hat n +C_N^\prime\bfg \sigma_N \hat n +
B_N(\bfg \sigma \hat {\bf k}) (\bfg \sigma_N \hat {\bf k})+\\ \nonumber
+ (G_N+H_N)(\bfg \sigma \hat {\bf q}) (\bfg \sigma_N \hat {\bf q})
+(G_N-H_N)(\bfg \sigma \hat {\bf n}) (\bfg \sigma_N \hat {\bf n}), 
\end{eqnarray}
where the complex numbers $A_N$, $C_N$, $C_N^\prime$, $B_N$, $G_N$, $H_N$ were fixed from 
the amplitudes of the SAID analysis \cite{Arndt:2007qn} and parametrized by a sum of Gaussians.
For the double scattering term in $pd$ scattering the unit vectors $\hat {\bf k}$, $\hat {\bf q}$,
$\hat {\bf n}$ are defined separately for each individual $NN$ collision.

The amplitude (\ref{pnamp}) is normalized in such a way that the invariant differential cross section 
has the following form:
 \begin{eqnarray}
\label{dsigpn} 
\frac{d\sigma_N}{dt}= \frac{1}{4} Tr M_NM_N^+.
\end{eqnarray}
 
Some additional results of calculations within this approach were reported recently in 
Ref.~\cite{Platonova:2019yzf}.

\begin{figure}[t]
\begin{center}
 \includegraphics[width=125mm]{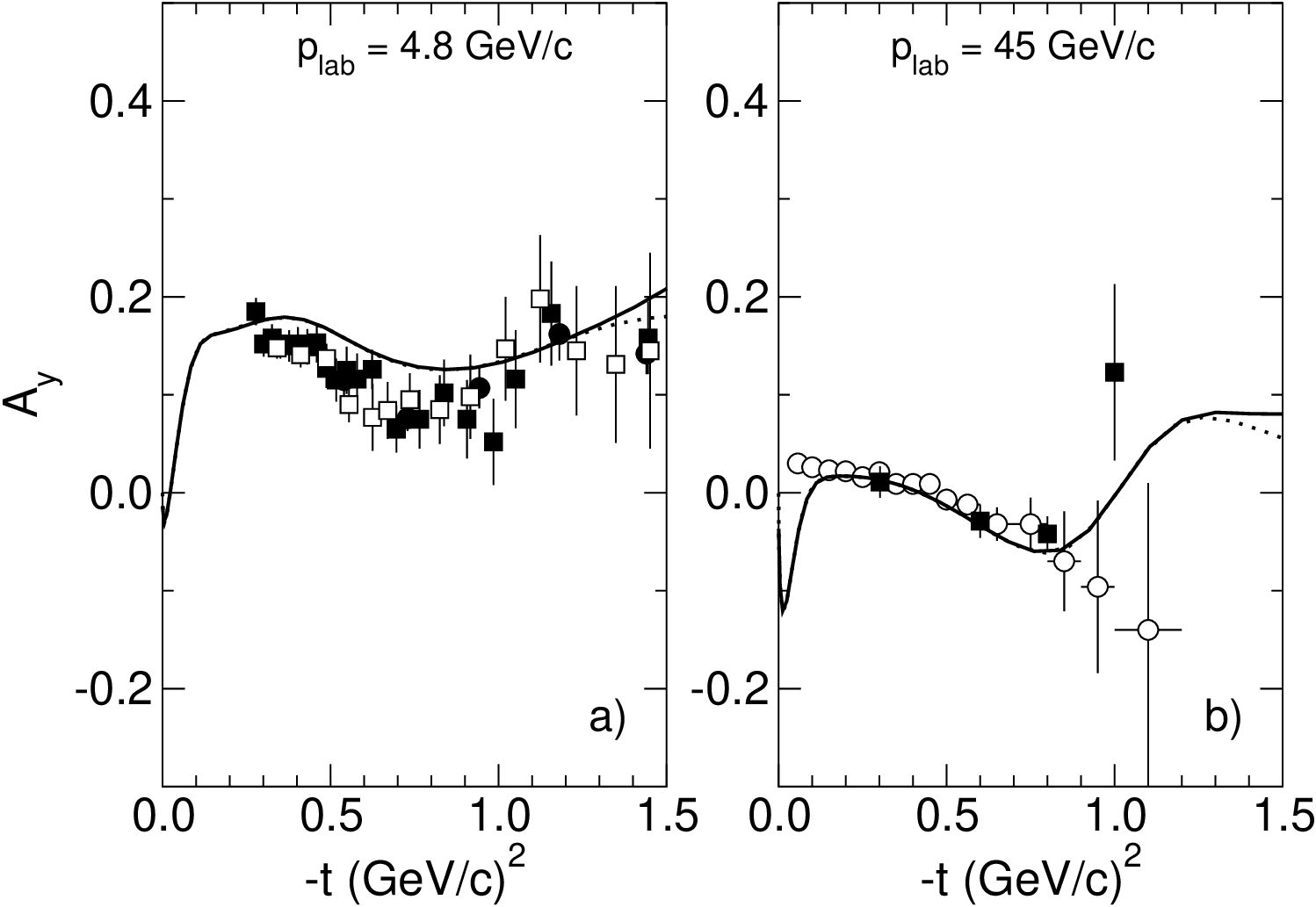} 
\vspace{-3mm}
\caption{Analyzing power for $pp$ elastic scattering as a function of
the four-momentum transfer $-t$ at $4.8$ GeV/c (left) and $45$ GeV/c (right). 
The dashed line is the result of the Regge model by Sibirtsev et al.~\cite{Sibirtsev:2009cz} 
while the solid line is based on the parameterization in terms of exponential 
functions (\ref{A-phi}). 
Left: Data are taken from Refs.~\cite{Parry:1973fj} (filled squares: $4.4$ GeV/c; 
open squares: $5.15$ GeV/c), and \cite{Abshire:1974ed} (circles). 
Right: Data are taken from Refs. \cite{Corcoran:1980ew} (squares)
and \cite{Gaidot:1976kj} (circles).  
}
\end{center}
\labelf{fig01}
\vspace{-5mm}
\end{figure}

\eject
\section*{Numerical results}

\begin{table} 
\caption{
Parameters of the $pN$ amplitudes at $p_{lab}=45$ GeV/c, cf.  Eq.~(\ref{A-phi}). 
The dimension $n$ for the coefficients $C_{\alpha,j}$ 
depends on the power of the $q$ factor in Eqs. (\ref{A-phi}) and is $n=1$
for $A_p$ and $G_p$, $n=2$ for $C_p$ and $n=3$ for $H_p$.  
}
\label{Tab1} 
\vskip 0.2cm
\renewcommand{\arraystretch}{1.2}
\begin{tabular} {|c|c|c|c|c|c|} \hline
 &   & \multicolumn{2}{|c|}{real part}& \multicolumn{2}{|c|}{imaginary part} \\
 & j & $C_{\alpha,j}$ (mb$^{1/2}$/GeV$^n$) & $A_{\alpha,j}$ (GeV$^{-2}$) 
& $C_{\alpha,j}$ (mb$^{1/2}$/GeV$^n$) & $A_{\alpha,j}$ (GeV$^{-2}$) \\ 
\hline 
\hline 
 & 1  & -0.10113182E+00 &  0.26000000E+01 & -0.50868385E-01 &  0.26000000E+01 \\
 & 2  & -0.30272788E+00 &  0.39000000E+01 &  0.45135636E+01 &  0.45000000E+01 \\
 & 3  &  0.36275621E-01 &  0.62769552E+01 &  0.54701498E+01 &  0.79740115E+01 \\
$A_p$  & 4  &  0.69044456E+00 &  0.93549981E+01 & -0.10151078E+02 &  0.12472690E+02 \\ 
 & 5  & -0.14278735E+01 &  0.13000000E+02 &  0.16289509E+02 &  0.17800000E+02 \\
 & 6  &  0.12152644E+01 &  0.17134442E+02 & -0.12268618E+02 &  0.23842646E+02 \\
 & 7  & -0.14350110E+01 &  0.21706020E+02 &  0.46423346E+01 &  0.30524183E+02 \\
\hline  
 & 1  &  0.74355080E-01 &  0.26000000E+01 &  0.45689189E-01 &  0.26000000E+01 \\
 & 2  & -0.16130899E+01 &  0.44000000E+01 & -0.79107203E+00 &  0.38000000E+01 \\
 & 3  &  0.45217973E+01 &  0.76911688E+01 &  0.57485752E+01 &  0.59941125E+01 \\
$C_p$  & 4  & -0.54780216E+01 &  0.11953074E+02 & -0.85442702E+01 &  0.88353828E+01 \\
 & 5  &  0.49658391E+01 &  0.17000000E+02 &  0.13994870E+02 &  0.12200000E+02 \\
 & 6  &  0.49231089E+01 &  0.22724612E+02 & -0.19190882E+02 &  0.16016408E+02 \\
 & 7  &  0.10015797E+02 &  0.29054489E+02 &  0.18732648E+02 &  0.20236326E+02 \\
\hline  
 & 1  & -0.74954883E+00 &  0.26000000E+01 & -0.11871787E+01 &  0.26000000E+01 \\
 & 2  &  0.40241315E+01 &  0.39000000E+01 &  0.53517954E+01 &  0.32000000E+01 \\
 & 3  & -0.12125361E+02 &  0.62769552E+01 & -0.96756963E+01 &  0.42970562E+01 \\
$G_p$  & 4  &  0.10072340E+02 &  0.93549981E+01 &  0.32522898E+01 &  0.57176913E+01 \\
 & 5  & -0.18247682E+01 &  0.13000000E+02 & -0.89637709E+01 &  0.73999998E+01 \\
 & 6  &  0.76091977E+00 &  0.17134442E+02 &  0.94782521E+01 &  0.93082036E+01 \\
 & 7  & -0.15772284E+00 &  0.21706020E+02 &  0.17443063E+01 &  0.11418163E+02 \\
\hline  
 & 1  &  0.31798006E+00 &  0.26000000E+01 &  0.43576664E+00 &  0.26000000E+01 \\
 & 2  & -0.33720909E+01 &  0.46000000E+01 & -0.25451131E+01 &  0.34000000E+01 \\
 & 3  &  0.24771736E+02 &  0.82568542E+01 &  0.42584276E+01 &  0.48627416E+01 \\
$H_p$  & 4  & -0.10758994E+02 &  0.12992305E+02 &  0.11014990E+02 &  0.67569218E+01 \\
 & 5  &  0.15394572E+02 &  0.18600000E+02 &  0.19721651E+02 &  0.89999998E+01 \\
 & 6  & -0.13147804E+02 &  0.24960680E+02 & -0.50146514E+00 &  0.11544272E+02 \\
 & 7  &  0.45696137E+01 &  0.31993877E+02 &  0.48863565E+00 &  0.14357550E+02 \\
\hline  
\end{tabular}
\renewcommand{\arraystretch}{1.0}
\end{table}

\begin{figure}[t]
\begin{center}
 \includegraphics[width=125mm]{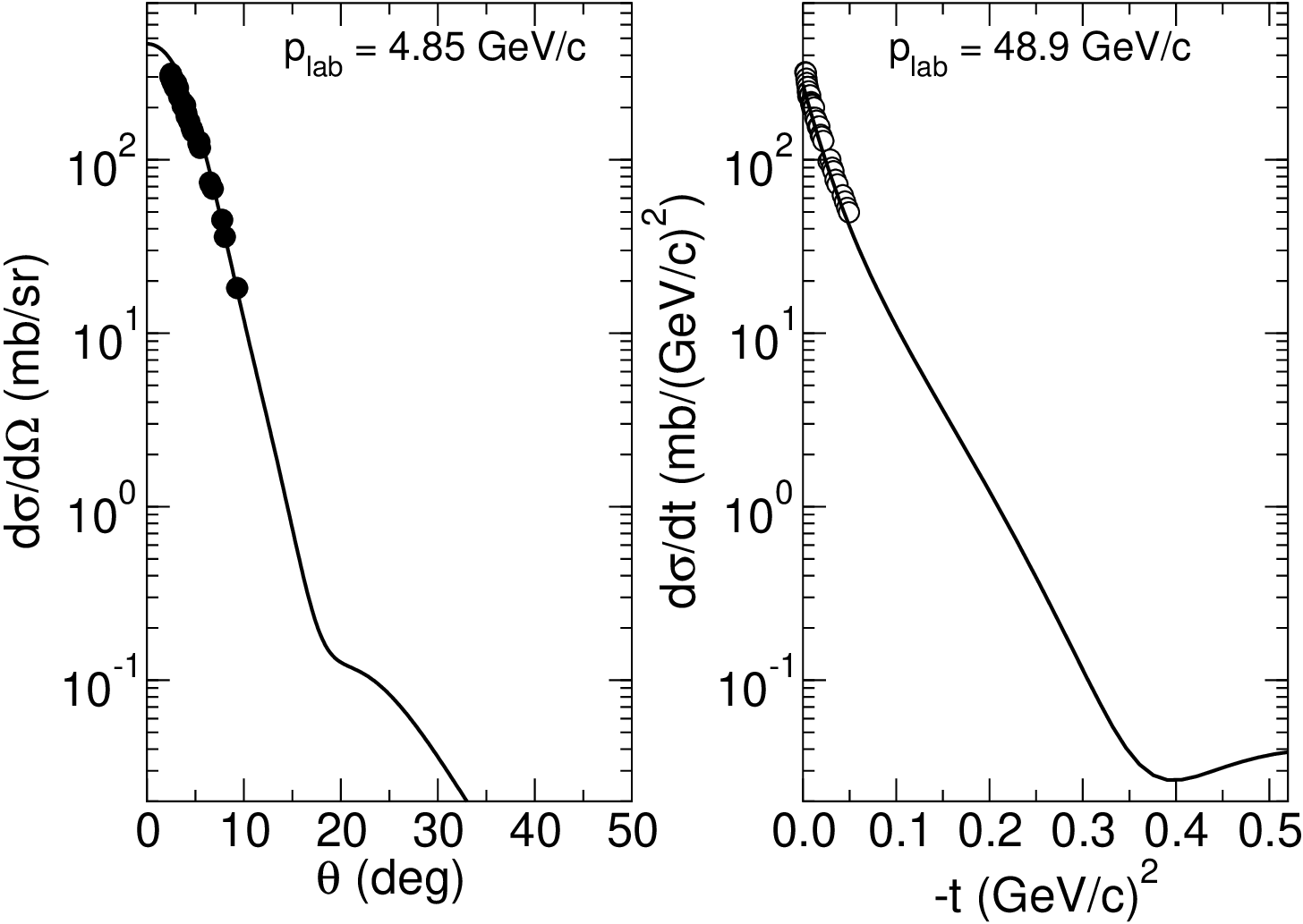}  
\vspace{-3mm}
\caption{Differential cross section for $pd$ elastic scattering.
Predictions are shown for $p_{lab} = 4.8$ GeV/c (left) and $45$ GeV/c
(right). 
Data are taken from Refs.~\protect{\cite{Dalkhazhav:1969cma}} ($4.8$ GeV/c) 
and \cite{Beznogikh:1973uka} ($48.9$ GeV/c). 
}
\end{center}
\labelf{fig02}
\vspace{-5mm}
\end{figure}

The relations between the $pN$ amplitudes $A_N$, $B_N$, $C_N$, $G_N$, $H_N$  
and the helicity amplitudes $\phi_1, \, ..., \, \phi_5$, 
and the corresponding expansion in terms of exponential functions are the following 
 \begin{eqnarray}
  \label{A-phi}
  A_N(q)=(\phi_3+\phi_1)/2 =\sum_i C_{a,j}\exp{(-A_{a,j}\,q^2)}, \nonumber \\
  B_N(q)= (\phi_3-\phi_1)/2=\sum_i C_{b,j}\exp{(-A_{b,j}\,q^2)}, \nonumber \\
  C_N(q)=i\phi_5=q\sum_i C_{c,j}\exp{(-A_{c,j}\,q^2)}, \nonumber \\
  G_N(q)=\phi_2/2=\sum_i C_{g,j}\exp{(-A_{g,j}\,q^2)}, \nonumber \\
  H_N(q)=\phi_4/2=q^2\sum_i C_{h,j}\exp{(-A_{h,j}\,q^2)},
 \end{eqnarray}
where $q^2=-t$. Note that $C_N'$ in Eq.~(\ref{pnamp}) is given by 
$C_N'(q) = C_N(q) + i\,(q/2m)\, A_N(q)$ \cite{Platonova:2016xjq}, with $m$ being the nucleon mass. 

Numerical values for the parameters of the Gaussians in Eqs. (\ref{A-phi}) are obtained by 
fitting to the helicity amplitudes from Ref.~\cite{Sibirtsev:2009cz}. Those for $p_{lab}=45$ GeV/c  
are summary in Table~\ref{Tab1}. Note that the parameters for the real and imaginary parts of the
amplitudes are given separately. 
Also, we would like to mention that $\phi_1 \equiv \phi_3$ 
in the model by Sibirtsev et al., see Eq.~(12) of that work,
and, accordingly, $B_N \equiv 0$. 
The differential cross section of $pp$ elastic scattering and the vector anayzing power 
$A_y$ are reproduced with these parameterizations on the same level of accuracy as in 
Ref.~\cite{Sibirtsev:2009cz}, in the interval of transferred four momentum 
$-t< 1.5$ (GeV/c)$^2$. As example, in Fig.~\ref{fig01} we show results for the latter, 
by the model (dotted line) and by the parameterization (solid line). 
 
In the next step we performed calculations for the $pd$ observables listed 
in Eqs.~(\ref{observs}) at $p_{lab}=4.68$ GeV/c and $45$ GeV/c, 
using these Gaussian parameterizations of the $pN$ amplitudes. 
Results for the differential cross section are shown in Fig.~\ref{fig02} while
a selection of spin-dependent observables is presented in Figs.~\ref{fig03} 
and \ref{fig04}. In the latter figures the results at $4.68$ GeV/c are 
indicated by the dashed lines, those at $45$ GeV/c by solid lines.  
One can see from Fig. \ref{fig02} that available data on the $pd$ elastic differential 
cross section in the forward hemisphere are well described by our calculations.
As was mentioned in the Introduction, we assume here that the $pn$ amplitudes are 
identical to the $pp$ amplitudes. Fig.~\ref{fig03} shows that the  
vector analyzing power $A_y^p$ decreases significantly in absolute value 
with increasing energy and a similar behaviour is exhibited by $A_y^d$.
In contrast, the spin correlation coefficients $C_{x,x}$ and $C_{y,y}$ show the 
opposite tendency, cf. Fig. \ref{fig04}. 
Coulomb effects are taken into account here in the 
$pp$ amplitude in the same way as in Ref.~\cite{Temerbayev:2015foa} 
and give a rather small contribution, as can be seen from Fig.~\ref{fig03}a,b).
One should note that the tensor analyzing powers $A_{xx}$ and $A_{yy}$, 
shown in Fig. \ref{fig04}, depend only weakly on the energy. Moreover, 
these observables do not change qualitatively in forward direction
if all spin-dependent amplitudes are excluded and only the spin-independent 
amplitude $A_N$ from Eq. (\ref{pnamp}) is taken into account, cf. the dotted
lines. On the hand, the spin correlation parameters $C_{x,x}$, $C_{y,y}$ 
practically vaninsh in this case (Fig. \ref{fig04}a,c)).

\begin{figure}[t]
\begin{center}
\includegraphics[width=125mm]{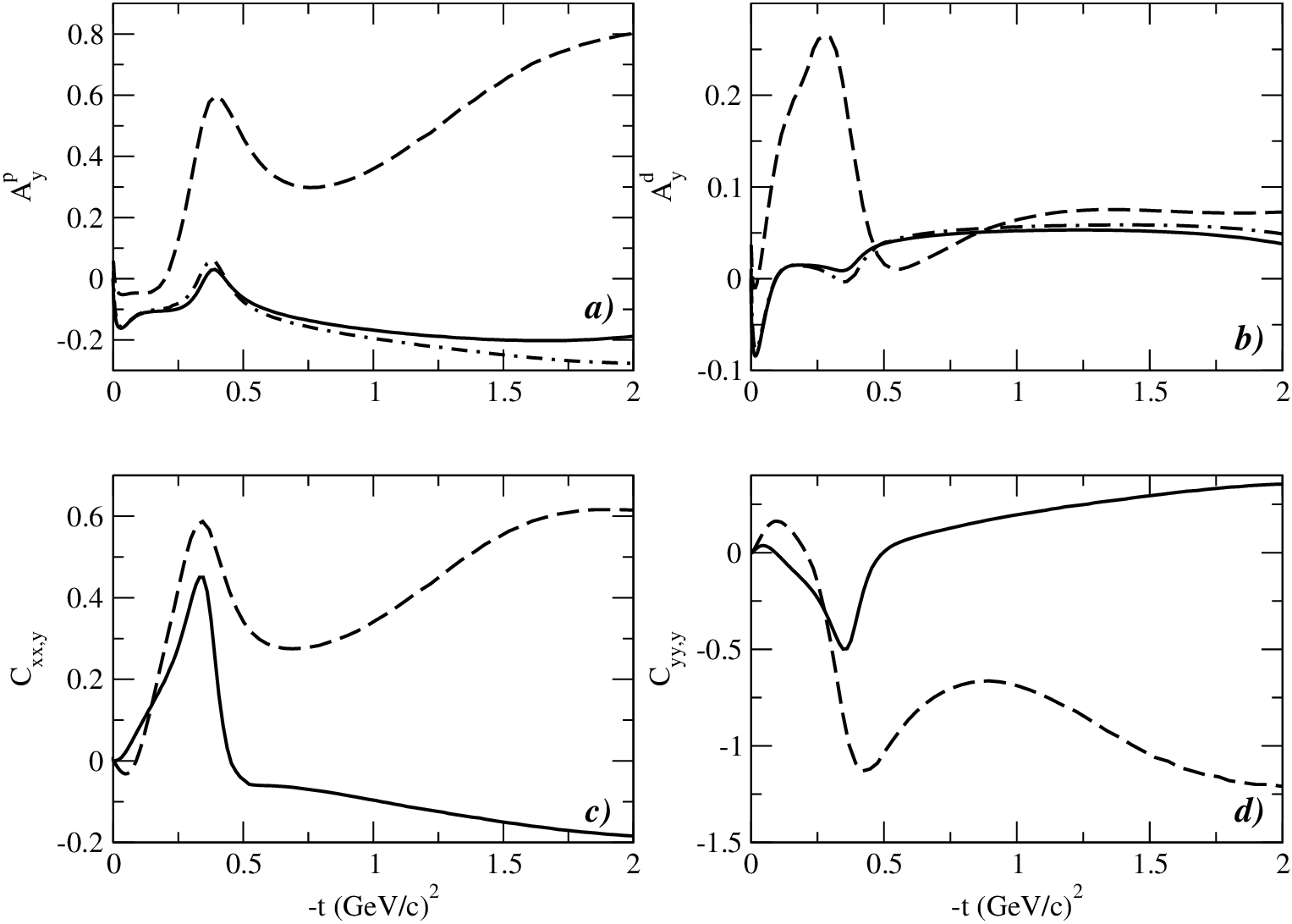}
\vspace{-2mm}
\caption{Results for spin-dependent $pd$ observables. 
Predictions for $p_{lab} = 4.8$ Gev/c are shown by dashed lines
while those at $45$ GeV/c correspond to the solid lines. 
The effect of the Coulomb interaction is indicated by the 
dash-dotted lines. 
}
\end{center}
\labelf{fig03}
\vspace{-5mm}
\end{figure}

\section*{Conclusion}

Nucleon-nucleon elastic scattering is a basic process in the physics of atomic 
nuclei and the interaction of hadrons with nuclei. Full information about the 
spin dependent $pN$ amplitudes can be obtained, in principle, from a complete 
polarization experiment, which, however, requires to measure twelve 
independent observables at a given collision energy and, thus, constitutes a 
too complicated experimental task. On the other hand existing models and 
corresponding parametrizations of $pp$ amplitudes in the region of small 
transferred momenta can be effectively tested by a measurement
of spin observables for $pd$ scattering and a subsequent comparison of 
the results with corresponding Glauber calculations. The spin observables
of $pd$ elastic scattering studied and evaluated in the 
\begin{figure}
\begin{center}
\includegraphics[width=125mm]{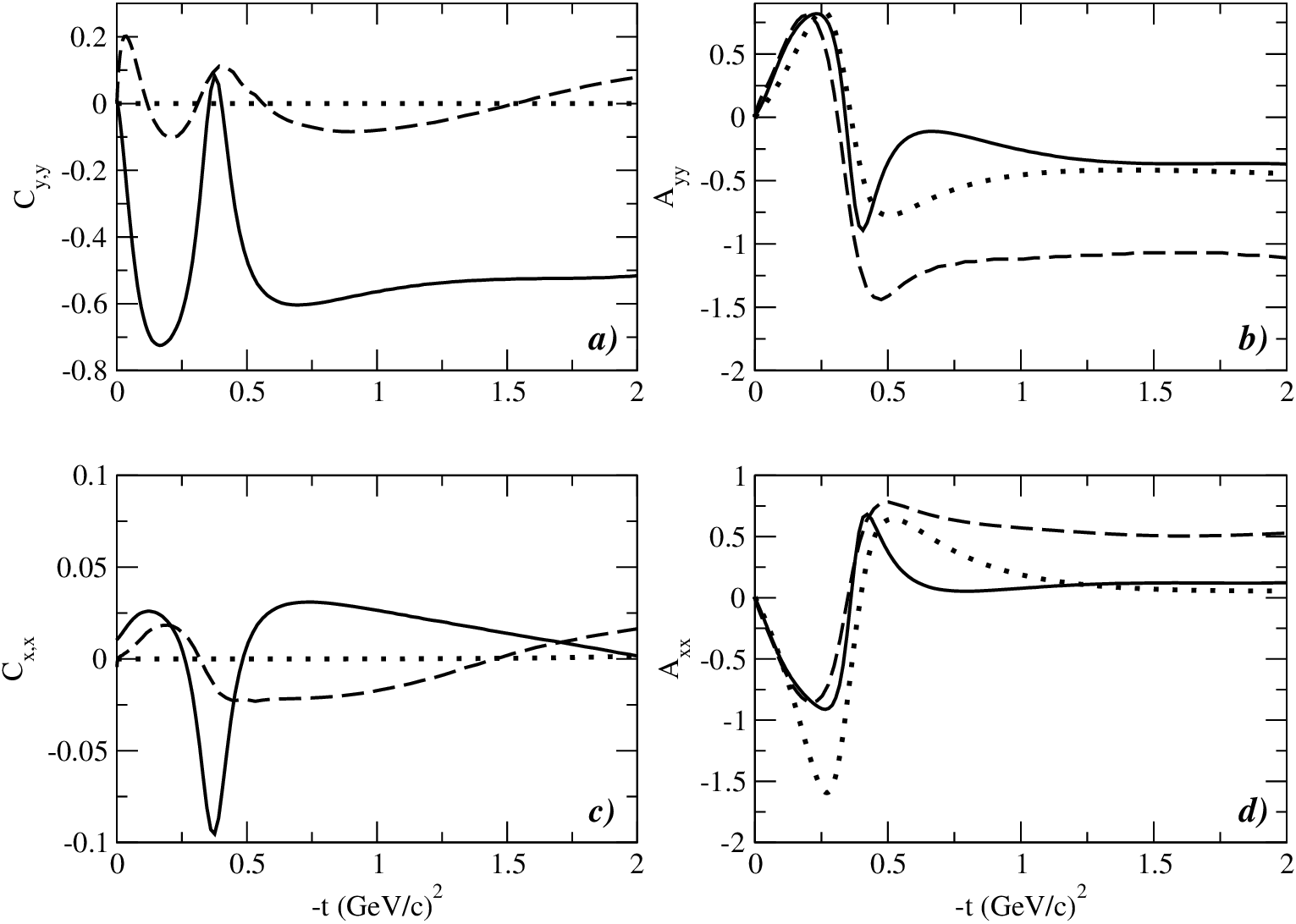} 
\vspace{-2mm}
\caption{Results for spin-dependent $pd$ observables. 
Same description of curves as in Fig.~\ref{fig03}.
The dotted lines are results where the spin-dependent $pN$
amplitudes have been omitted in the calculation. 
}
\end{center}
\labelf{fig04}
\vspace{-5mm}
\end{figure}
present work 
are found to be not too small and, thus, could be measured at the future 
SPD NICA facility. 

\vskip 0.5cm
 {\bf Acknowledgements:} A.B., A.T. and Yu.U. acknowledge their support
  by  the Scientific Programme of JINR-Kazakhstan  N 391-20.07.2020. 
\vskip 2cm

\nocite{*}
\bibliographystyle{pepan}
%

 \end{document}